\documentclass[twocolumn]{aastex631}

\usepackage{url}
\usepackage{amsmath}
\usepackage{booktabs}
\usepackage{graphicx}
\usepackage{hyperref}

\begin{document}

\title{MAMMOTH-Subaru V. Effects of Cosmic Variance on Ly$\alpha$ Luminosity Functions at $z=2.2-2.3$}

\author{Ke Ma}
\affiliation{Department of Astronomy, Tsinghua University, Beijing 100084, People’s Republic of China; \url{markma501x@gmail.com};\url{haibinzhang@mail.tsinghua.edu.cn}}

\author{Haibin Zhang}
\affiliation{Department of Astronomy, Tsinghua University, Beijing 100084, People’s Republic of China; \url{markma501x@gmail.com};\url{haibinzhang@mail.tsinghua.edu.cn}}

\author{Zheng Cai}
\affiliation{Department of Astronomy, Tsinghua University, Beijing 100084, People’s Republic of China; \url{markma501x@gmail.com};\url{haibinzhang@mail.tsinghua.edu.cn}}

\author{Yongming Liang}
\affiliation{Institute for Cosmic Ray Research, The University of Tokyo, Kashiwa, Chiba 277-8582, Japan}

\author{Nobunari Kashikawa}
\affiliation{National Astronomical Observatory of Japan, 2-21-1 Osawa, Mitaka, Tokyo 181-8588, Japan}
\affiliation{Department of Astronomy, School of Science, The University of Tokyo, 7-3-1 Hongo, Bunkyo-ku, Tokyo 113-0033, Japan}
\affiliation{Research Center for the Early Universe, The University of Tokyo, 7-3-1 Hongo, Bunkyo-ku, Tokyo 113-0033, Japan}

\author{Mingyu Li}
\affiliation{Department of Astronomy, Tsinghua University, Beijing 100084, People’s Republic of China; \url{markma501x@gmail.com};\url{haibinzhang@mail.tsinghua.edu.cn}}

\author{Yunjing Wu}
\affiliation{Department of Astronomy, Tsinghua University, Beijing 100084, People’s Republic of China; \url{markma501x@gmail.com};\url{haibinzhang@mail.tsinghua.edu.cn}}
\affiliation{Steward Observatory, University of Arizona, 933 N Cherry Ave, Tucson, AZ 85721, USA}

\author{Qiong Li}
\affiliation{Jodrell Bank Centre for Astrophysics, University of Manchester, Oxford Road, Manchester, UK}

\author{Xiaohui Fan}
\affiliation{Steward Observatory, University of Arizona, 933 N Cherry Ave, Tucson, AZ 85721, USA}

\author{Sean D. Johnson}
\affiliation{Department of Astronomy, University of Michigan, Ann Arbor, MI 48109, USA}

\author{Masami Ouchi}
\affiliation{National Astronomical Observatory of Japan, 2-21-1 Osawa, Mitaka, Tokyo 181-8588, Japan}
\affiliation{Institute for Cosmic Ray Research, The University of Tokyo, Kashiwa, Chiba 277-8582, Japan}
\affiliation{Kavli Institute for the Physics and Mathematics of the Universe (WPI), University of Tokyo, Kashiwa, Chiba 277-8583, Japan}


\begin{abstract}
    Cosmic variance introduces significant uncertainties into galaxy number density properties when surveying the  high-redshift Universe with a small volume, such uncertainties produce the field-to-field variance of galaxy number $\sigma_{g}$ in observational astronomy. 
    This uncertainty significantly affects the Luminosity Functions (LF) measurement of Lyman-alpha (Ly$\alpha$) Emitters (LAEs).
    For most previous Ly$\alpha$ LF studies, $\sigma_{g}$ is often adopted from predictions by cosmological simulations, but barely confirmed by observations. 
    Measuring cosmic variance requires a huge sample over a large volume, exceeding the capabilities of most astronomical instruments. 
    In this study, we demonstrate an observational approach for measuring the cosmic variance contribution for $z\approx2.2$ Ly$\alpha$ LFs. 
    The Ly$\alpha$ emitter candidates are observed using narrowband and broadband of the Subaru/Hyper Suprime-Cam (HSC), with 8 independent fields, making the total survey area $\simeq$ 11.62 deg$^2$ and a comoving volume of $\simeq$ 8.71 $\times$ 10$^6$ Mpc$^3$. 
    These eight fields are selected using the project of MApping the Most Massive Overdensity Through Hydrogen (MAMMOTH) \cite[e.g.][]{Cai_etal_2017a}. 
    We report a best-fit Schechter function with parameters $\alpha=-1.75$ (fixed), $L_{Ly\alpha}^{*}=5.18_{-0.40}^{+0.43}$ $ \times 10^{42}$ erg s$^{-1}$ and $\phi_{Ly\alpha}^{*}=4.87_{-0.55}^{+0.54}$ $ \times 10^{-4}$ Mpc$^{-3}$ for the overall Ly$\alpha$ LFs.
    After clipping out the regions that can bias the cosmic variance measurements, we calculate the field-to-field variance $\sigma_{g}$, by sampling LAEs within multiple pointings assigned on the field image. 
    We investigate the relation between $\sigma_{g}$ and survey volume $V$, and fit a simple power law: $\sigma_g = k \times (\frac{V_{\rm eff}}{10^5 {\rm Mpc}^3})^{\beta}$.
    We find best-fit values of $-1.209_{-0.106}^{+0.106}$ for $\beta$ and $0.986_{-0.100}^{+0.108}$ for k. We compare our measurements with predictions from simulations and find that the cosmic variance of LAEs might be larger than that of general star-forming galaxies.
\end{abstract}


\section{INTRODUCTION}

The cosmic structure formation arises from fluctuations during the inflationary epoch \cite[][]{Guth_1981} and introduces inhomogeneity to the Universe except on large scales.
Due to the presence of the scale-dependent inhomogeneity, observations on the number density and Luminosity Function (LF) within a small survey volume have an inevitable level of uncertainty \cite[e.g.][]{Szapudi_Colombi_1996,Trenti_etal_2008,Robertson_2010b}, known as the cosmic variance. 
Note cosmic variance is referred to as a sample variance due to different survey sizes within our Universe, instead of arising from an ensemble across multiple Universes. 
This source of error dominates uncertainties when the survey volume is smaller than the typical clustering scale, or at high redshift where galaxies tend to be more clustered than dark matter \cite[][]{Kauffmann_etal_1999,Coil_etal_2004}. 
Therefore, cosmic variance is significant in observational studies of luminosity functions (LF) for high redshift galaxies, where the number density is expressed as a function of the emission line luminosity \cite[For a detailed approach, see][]{Gronke_etal_2015}. 

Ly$\alpha$ emission lines are one of the most searched and surveyed for studying high redshift galaxies. 
Star-forming galaxies with a strong Ly$\alpha$ emission line, referred to as Ly$\alpha$ Emitters (LAEs), have been widely identified through deep narrowband imaging and spectroscopic surveys \cite[e.g.][]{Cowie_Hu_1998,Ouchi_etal_2008,Hayes_etal_2010,Konno_etal_2016,Itoh_etal_2018,Shibuya_etal_2019}. 
It is commonly agreed that Ly$\alpha$ LFs show a dramatic increase from $z\sim 0$ to $z\sim 3$  \cite[e.g.][]{Deharveng_etal_2008,Cowie_etal_2010,Cowie_etal_2011,Barger_etal_2012}, fewer evolutions from $z\simeq 3$ to $z \simeq 6$   \cite[e.g.][]{Gronwall_etal_2007,Ouchi_etal_2008}, 
and a decrease beyond $z\approx 6$  \cite[e.g.][]{Kashikawa_etal_2006,Ouchi_etal_2010,Konno_etal_2018,Konno_etal_2014,Itoh_etal_2018}. 

Although many works have been done for LAE surveys at $z\sim2-3$ \cite[e.g.][]{Matthee_etal_2017,Sobral_etal_2018,Spinoso_etal_2020}, significant disagreements existed between different studies (e.g. \cite{Ciardullo_etal_2012} and e.g. \cite{Konno_etal_2016}), and it has been suggested by many studies \cite[e.g.][]{Sobral_etal_2015,Sobral_etal_2017} that strong cosmic variance can be a possible explanation. 
Thus, quantifying the cosmic variance is required to understand the formation of the cosmic structures. 

The fluctuations in the galaxy number densities caused by cosmic variance are sometimes referred to as the field-to-field variations and are often denoted as $\sigma_g$ where $g$ stands for galaxies.
The value of this $\sigma_g$ can vary with galaxy types, redshift, survey volumes, and sometimes the survey geometry \cite[e.g.][]{Newman_Davis_2002,Moster_etal_2010,Robertson_2010a}. 
Theoretically, the field-to-field variation for a certain type of galaxy with a similar survey geometry (e.g. `pencil beam' shape for narrowband imaging surveys) at a given redshift only depends on the survey volume. 
In general, measuring the field-to-field variance requires a vast sampling, however, due to the relatively small sample size, it is difficult to estimate the cosmic variance with a sufficient level of accuracy for previous studies. 
Therefore, the common approach is to use the field-to-field variation predicted from cosmological simulations \cite[e.g.][]{Newman_Davis_2002,Trenti_etal_2008,Moster_etal_2010}. 
Nevertheless, such a field-to-field variance has not been confirmed observationally. 

In this study, we provide an observational approach for investigating the cosmic variance through Ly$\alpha$ LFs of $z\sim 2.2$ LAEs. 
We have developed a simple relation between $\sigma_g$ and the survey volume for $z\sim 2.2$ narrowband imaging surveys. 
The data are obtained from the Subaru/Hyper Suprime-Cam (HSC, \cite{Miyazaki_etal_2018}) with a wide Field-of-View (FoV) of 1.5 deg in diameter, which gives us a large survey volume available for Monte-Carlo sampling. 
Eight fields targeting the candidates of the MApping the Most Massive Overdensity Through Hydrogen (MAMMOTH) project \cite[][]{Cai_etal_2016,Cai_etal_2017a,Cai_etal_2017b} are used, with each field covered by one HSC FoV. 

This paper is organized as follows: 
We summarised the information on observational data and target fields, the data processing, and the selection criteria of LAEs in Section 2. 
A detailed explanation for computing the Ly$\alpha$ LFs is described in Section 3. 
In Section 4, we demonstrate the method and results for the field-to-field variance estimations using Ly$\alpha$ LFs. 
Finally, a summary of this study is listed in Section 5. 
Throughout this paper, we adopt the AB magnitude system \cite[][]{Oke_Gunn_1983} and a flat $\Lambda$CDM cosmology with $\Omega_{\rm m} = 0.3 $, $\Omega_{\rm \Lambda}= 0.7 $, and $h = 0.70$.

\section{DATA}
\subsection{Targeting fields and observations}

The fields used for this study are selected using the background quasar spectra from the Baron Oscillation Spectroscopic Survey (BOSS) of SDSS \cite{Dawson_etal_2013,Dawson_etal_2016}. 
We target a total number of 8 overdense fields centered at MAMMOTH candidates, with two narrowband filters covering 4 fields each. 
For the NB387 filter, the fields covered are BOSS$J0210+0052$ (J0210), BOSS$J0222-0224$ (J0222), BOSS$J0924+1503$ (J0924), and BOSS$J1419+0500$ (J1419). 
For the NB400 filter, the fields covered are BOSS$J0240-0521$ (J0240), BOSS$J0755+3108$ (J0755), BOSS$J1133+1005$ (J1133), and BOSS$J1349+2427$ (J1349). 
The relevant information of 8 fields is summarised in table \ref{tab:fdinfo}. See also \cite{Liang_etal_2021} and Cai et al. in prep. for details of field selections.

\begin{figure}[ht]
    \includegraphics[width=8.5cm]{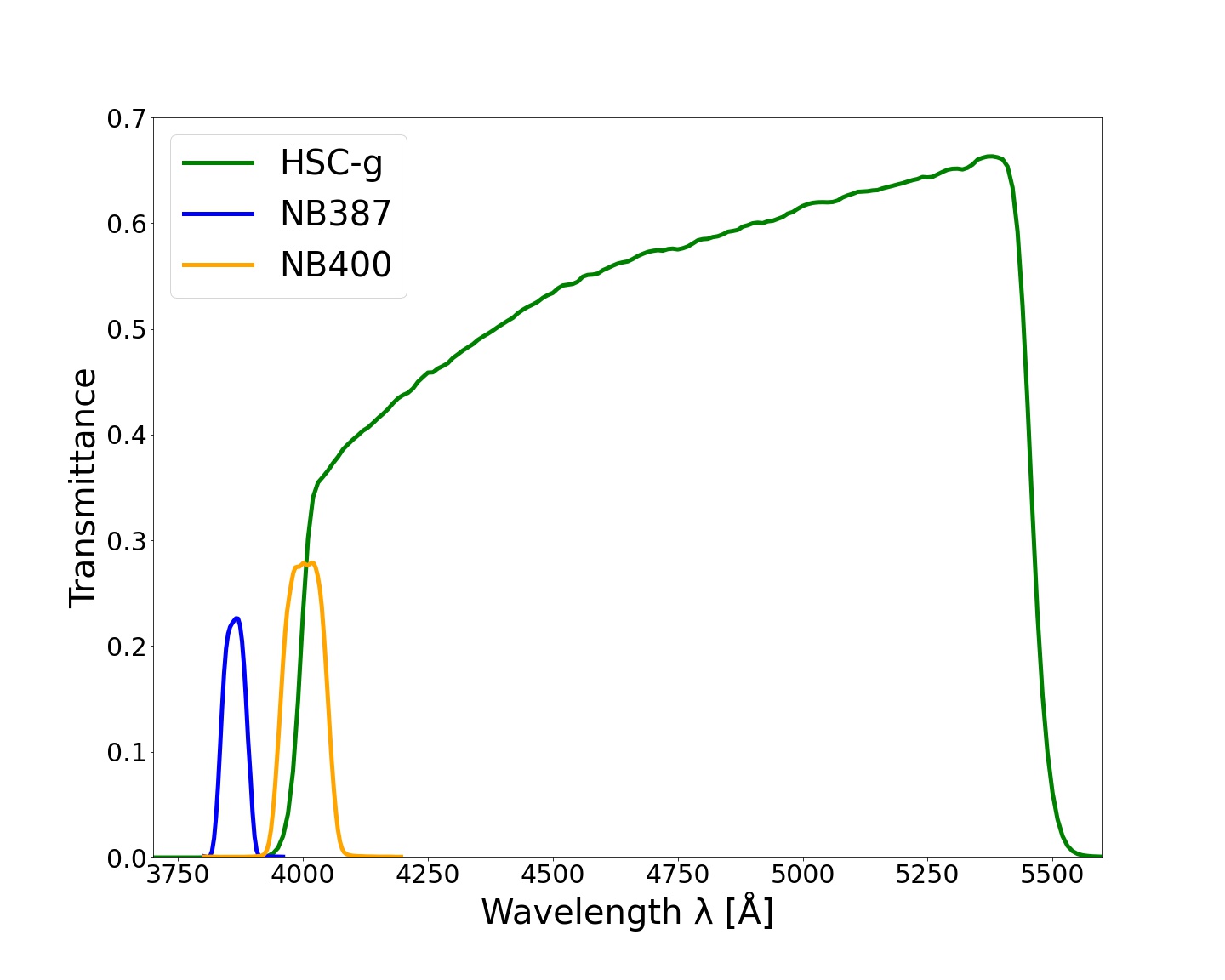}
    \caption{Transmission curve of the NB387 (blue), NB400 (yellow) and HSC-g (green) band. The curves represent the total transmittance accounting in CCD quantum efficiency, dewar window, the Primary Focus Unit, and the reflectivity of the Prime Mirror.}
    \label{fig:transM}
\end{figure}

We obtained our observations using the gigantic mosaic CCD camera HSC, with a wide Field-of-View (FoV) of 1.5 deg in diameter, and a pixel scale of 0.168''. 
Narrowband imagings are carried out using two NB filters, NB387 ($\lambda_c$ = 3862 Å, FWHM = 56 Å) and NB400 ($\lambda_0$ = 4003 Å, FWHM = 92 Å). 
The corresponding redshifts of detecting Ly$\alpha$ emissions are $z = 2.18 \pm 0.02$ and $z = 2.29 \pm 0.04$ respectively. 
We also use the g-band to evaluate the continuum of the detected objects. 
Figure \ref{fig:transM} shows the transmission curve of these 3 filters used, which takes the transmittance accounting in CCD quantum efficiency, dewar window, and the Primary Focus Unit and the reflectivity of the Prime Mirror into account. 
The imaging data are reduced with the HSC pipeline, $hscPipe$ \cite[][]{Bosch_etal_2018,Aihara_etal_2019}, see \cite{Liang_etal_2021} and Zhang et al. in prep. for details of data reduction. 
Due to the poor quality of the J0210 NB387 data, NB387 and g-band data of J0210 field are reduced with two different versions of $hscPipe$, which results in a slight difference in the final image depth, as shown in table \ref{tab:fdinfo}. 
The final catalog of J0210 is a combination of the two. 

\begin{table}[h!]
\centering
\begin{tabular}{c c c c c c} 
 \hline
 Field & J2000 RA & J2000 DEC & A$_{\rm eff}$ & m$_{{\rm NB},5 \sigma}$ & m$_{{\rm g},5 \sigma}$ \\ [0.5ex] 
 name & hh:mm:ss & dd:mm:ss & deg$^2$ & mag & mag \\ [0.5ex]
 \hline\hline
 J0210 & 02:09:58.90 & +00:53:43.0 & 1.34 & 24.36 & 26.24  \\ [1ex] 
       &             &             &      & 24.25 & 26.34  \\ [1ex] 
 J0222 & 02:22:24.66 & $-$02:23:41.2 & 1.13 & 24.99 & 27.01  \\ [1ex] 
 J0924 & 09:24:00.70 & +15:04:16.7 & 1.47 & 24.74 & 26.63  \\ [1ex]
 J1419 & 14:19:33.80 & +05:00:17.2 & 1.45 & 24.81 & 26.80  \\ [1ex] 
 J0240 & 02:40:05.11 & $-$05:21:06.7 & 1.53 & 25.61 & 26.80  \\ [1ex]
 J0755 & 07:55:35.89 & +31:09:56.9 & 1.54 & 25.83 & 26.50  \\ [1ex] 
 J1133 & 11:33:02.40 & +10:05:06.0 & 1.55 & 25.39 & 26.30  \\ [1ex] 
 J1349 & 13:49:40.80 & +24:28:48.0 & 1.62 & 25.67 & 26.15  \\ [1ex] 
 \hline
\end{tabular}
\caption{Summary of field information. Column 1 is the name of fields; Columns 2 and 3 are the RA and DEC coordinates in equinox with an epoch of J2000; Column 4 is the effective survey area after applying masks; Columns 5 and 6 are the 5$\sigma$ limiting magnitudes measured within a 1.7'' (2.5'' for J0210 image) aperture for the final stacked NB387/NB400 image and the PSF-matched g-band, respectively. Data of J0210 field is reduced using two versions of $hscPipe$, whose depths are also indicated in the first two rows.}
\label{tab:fdinfo}
\end{table}

\subsection{Photometric processing}

Object detection and the following photometry are performed using the $SExtractor$ \cite{Bertin_Arnouts_1996}. 
The g-band and NB387/400 images are PSF matched by convolving a proper Gaussian kernel in each field (\cite{Liang_etal_2021}, Zhang et al. in prep.). 
The detection threshold is set as 15 continuous pixels over the 1.2$\sigma$ sky background. 
We also apply masks to the pixels with low S/N signals or saturated by bright stars and artifacts during object detection. 
In addition, the sky background root-mean-square map is used as the weighting map in $SExtractor$, to minimize the influence of the fluctuations in the image depth for each field. 
The aperture diameters for photometry are 15 pixels ($\sim$ 2.5'') for J0210 field and 10 pixels ($\sim$ 1.7'') for the other 7 fields.

We estimate the total magnitude using AUTO-MAG, which uses the automatically determined elliptical aperture for Kron photometry by $SExtractor$ \cite{Bernardeau_Kofman_1995}. 
Both AUTO-MAG and APER-MAG output by $SExtractor$ are used in the selection criteria described in the next section. 
Furthermore, we also replace the g-band magnitudes with the corresponding 2$\sigma$ limiting magnitudes, for the objects with a g-band fainter than the 2$\sigma$ limit. 
The 5$\sigma$ limiting magnitudes within the 1.7'' (2.5'' for J0210) aperture of the final stacked g-band and NB387/400 images are listed in table \ref{tab:fdinfo}. 
The depths for the NB400 images are about 1 mag deeper than that for the NB387 images in general, the seeing and the depth of J0210 are relatively poorer than all other fields.

\subsection{LAE Selection}

We use the narrowband color excesses in the broadband to select out our LAE samples, this approach is widely used in previous studies \cite[e.g.][]{Guaita_etal_2010,Ouchi_etal_2008,Konno_etal_2016}. 
Most of these studies use multi-wavelength data and thus multiple color excesses to define their selection criteria (e.g. U-band), while we only use one broadband (HSC g-band) to estimate the continuum. 
However, \cite{Liang_etal_2021} have proved that it is sufficient enough for the z$\sim 2.2$ LAE selection. 

\begin{figure}[ht]
    \includegraphics[width=8.5cm]{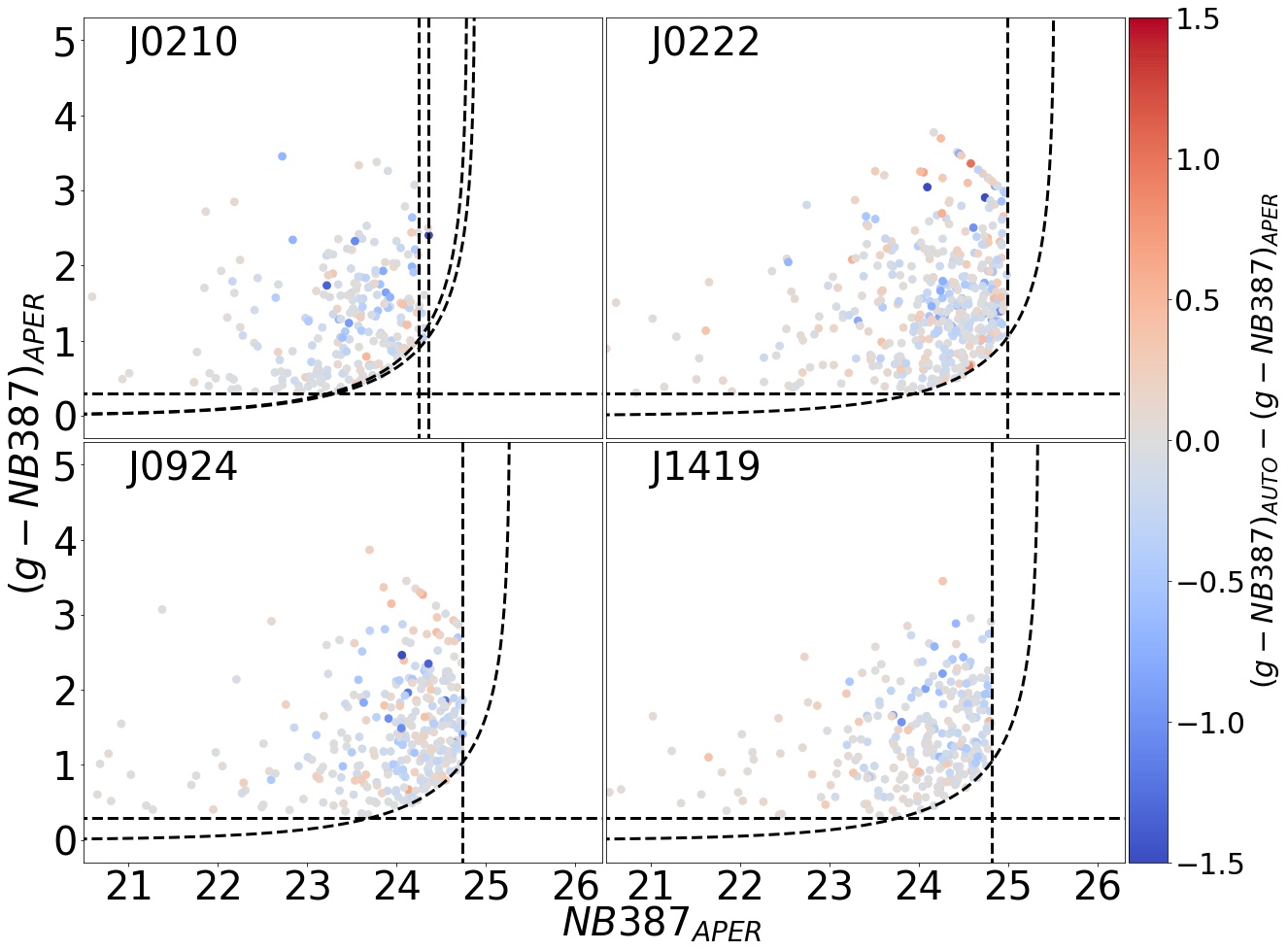}
    \vfill
    \includegraphics[width=8.5cm]{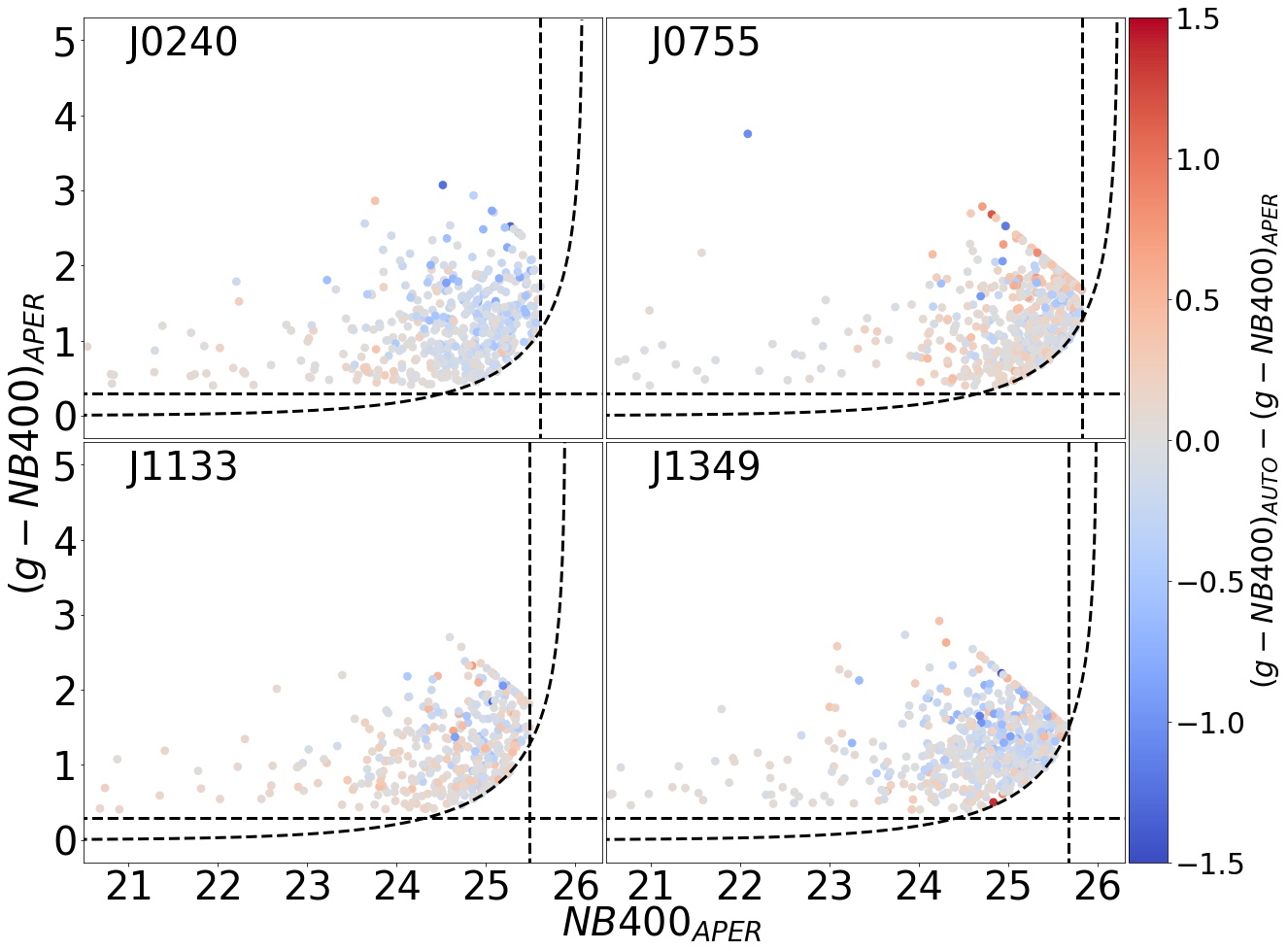}
    \caption{g - NB vs. NB diagram of APER-MAG for LAE selections in each field. The color bar shows the the difference between the g - NB values calculated from AUTO-MAG and that from APER-MAG. The first three selection criteria for the APER-MAGs are plotted as black dashed lines. For J0210, two versions of data reductions yield two different image depths, thus both criteria are overplotted for clarification. Furthermore, the g-band magnitudes are replaced as the corresponding 2$\sigma$ limiting magnitude, for any object with a g-band fainter than the respective 2$\sigma$ limit. The g - NB values for these objects are shown as a lower limit in the diagrams.}
    \label{fig:seles}
\end{figure}

We assume the LAE spectrum has a Gaussian-like redshifted Ly$\alpha$ emission with a rest Equivalent Width (EW) of 20 Å, and a continuum that follows a simple power law of $f_{\lambda} = \lambda^{\beta}$. 
We also take IGM absorption \cite[][]{Madau_1995,Madau_Dickinson_2014} into account when calculating the observed magnitude \cite[][]{Inoue_etal_2014}. 
In addition, we find that the value of g - NB calculated from APER-MAG is not necessarily the same as that from AUTO-MAG, as shown in figure \ref{fig:seles} (The differences between g-NB values calculated from AUTO-MAG and that from APER-MAG are not zero for every object). 
As a result, for some objects with APER-MAG values passing the first three criteria, its g - NB value calculated from AUTO-MAG can still be smaller than the corresponding EW$_0$ = 20 \AA\ values, or even smaller than 0. 
Since we use AUTO-MAG to estimate the total magnitude, this could be a problem when calculating the Ly$\alpha$ luminosity. 
To address this issue, we also apply the criteria for assessing the color excess and color error to the AUTO-MAG, so that (g - NB)$_{\rm AUTO}$ is also above the required values. \par
\vspace{1em}
\noindent For NB387 data, the selection criteria are defined as:
\begin{equation}
    \begin{aligned}
        & 20.5 < {\rm NB}387_{\rm APER} \lesssim {\rm NB}_{{\rm lim},5 \sigma} \\
        & (g - {\rm NB}387)_{\rm APER} > 0.3 \\
        & (g - {\rm NB}387)_{\rm APER} > 3 \sigma({\rm NB}387_{\rm APER}) \\
        & (g - {\rm NB}387)_{\rm AUTO} > 0.3 \\
        & (g - {\rm NB}387)_{\rm AUTO} > 3 \sigma({\rm NB}387_{\rm AUTO}) \\
    \end{aligned}
    \label{eqn:select387}
\end{equation}
For NB400 data, the selection criteria are defined as:
\begin{equation}
    \begin{aligned}
        & 18.0 < {\rm NB}400_{\rm APER} \lesssim {\rm NB}_{{\rm lim},5 \sigma} \\
        & (g - {\rm NB}400)_{\rm APER} > 0.4 \\
        & (g - {\rm NB}400)_{\rm APER} > 3 \sigma({\rm NB}400_{\rm APER}) \\
        & (g - {\rm NB}400)_{\rm AUTO} > 0.4 \\
        & (g - {\rm NB}400)_{\rm AUTO} > 3 \sigma({\rm NB}400_{\rm AUTO}) \\
    \end{aligned}
    \label{eqn:select400}
\end{equation}
Here, The first line defines the detection limits using APER-MAG, where the lower limits  are set to avoid saturations, and the upper limits are set as the corresponding $5 \sigma$ limiting magnitudes to ensure the reliability of the source detections. 
The color excesses are equivalent to setting EW$_0 >$ 20 Å, which are $\sim 0.3$ for NB387 filter and $\sim 0.4$ for NB400 filter. 
$3 \sigma$(NB) corresponds to the color error, and is defined following \cite{Shibuya_etal_2018}:
\begin{equation}
    3 \sigma({\rm NB}) = -2.5 {\rm log} (1 - 3 \times \frac{\sqrt{f_{1 \sigma, {\rm NB}}^2+f_{1 \sigma, {\rm g}}^2}}{f_{\rm NB}})
    \label{eqn:colorerror}
\end{equation}
which is designed to reject false selections of faint contaminants passing the criteria due to statistical fluctuations. 
We plot the LAE selections for each field in figure \ref{fig:seles}. 
Moreover, the NB387 data are re-selected from the original catalog of \cite{Liang_etal_2021}, where a 2$\sigma$ color error and APER-MAG-only criteria are used. 
We make the selection criteria consistent across two filters in this study to investigate the field-to-field variations. 

\section{LUMINOSITY FUNCTION}
\subsection{Detection completeness}

\begin{figure}[ht]
    \includegraphics[width=8.5cm]{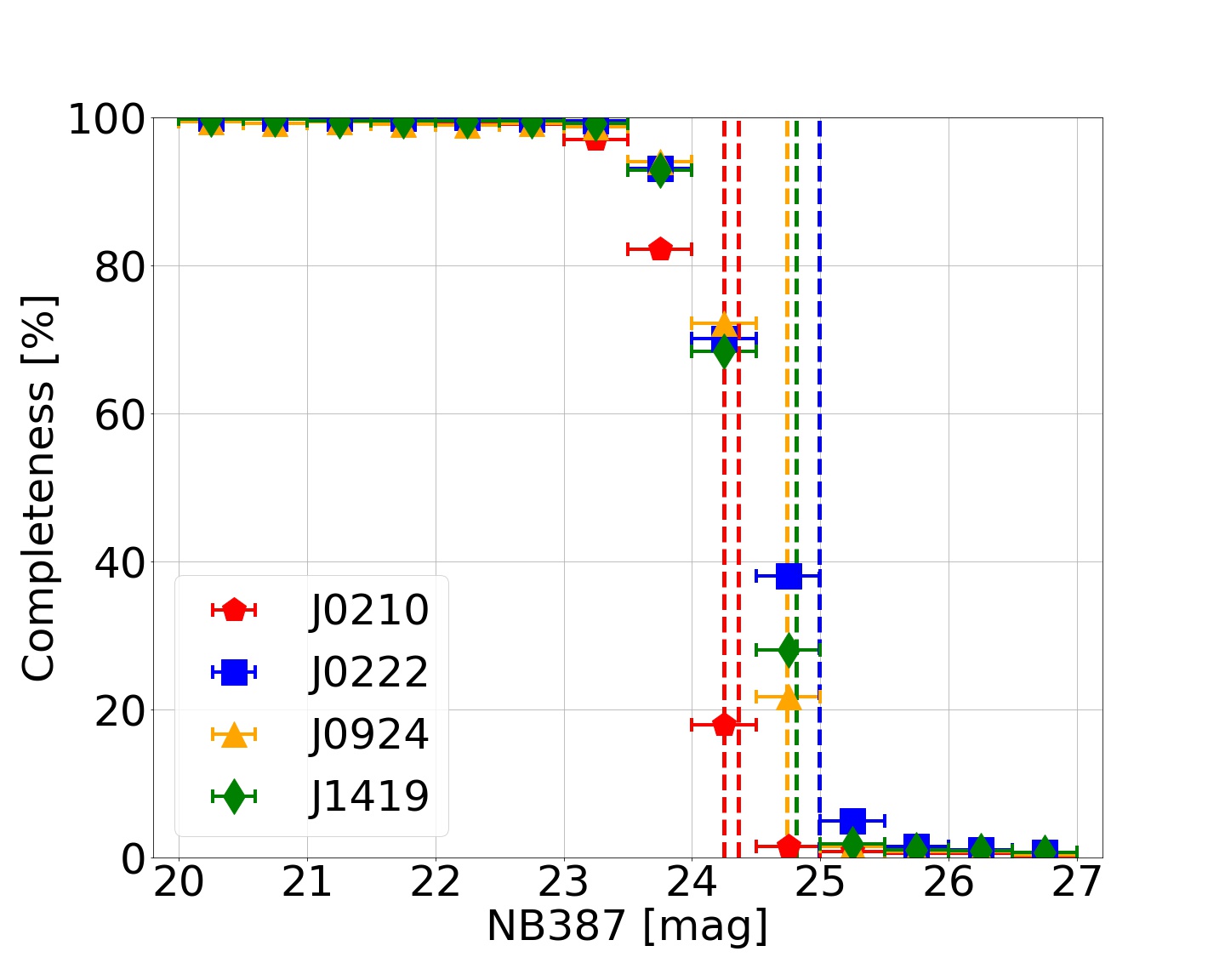}
    \vfill
    \includegraphics[width=8.5cm]{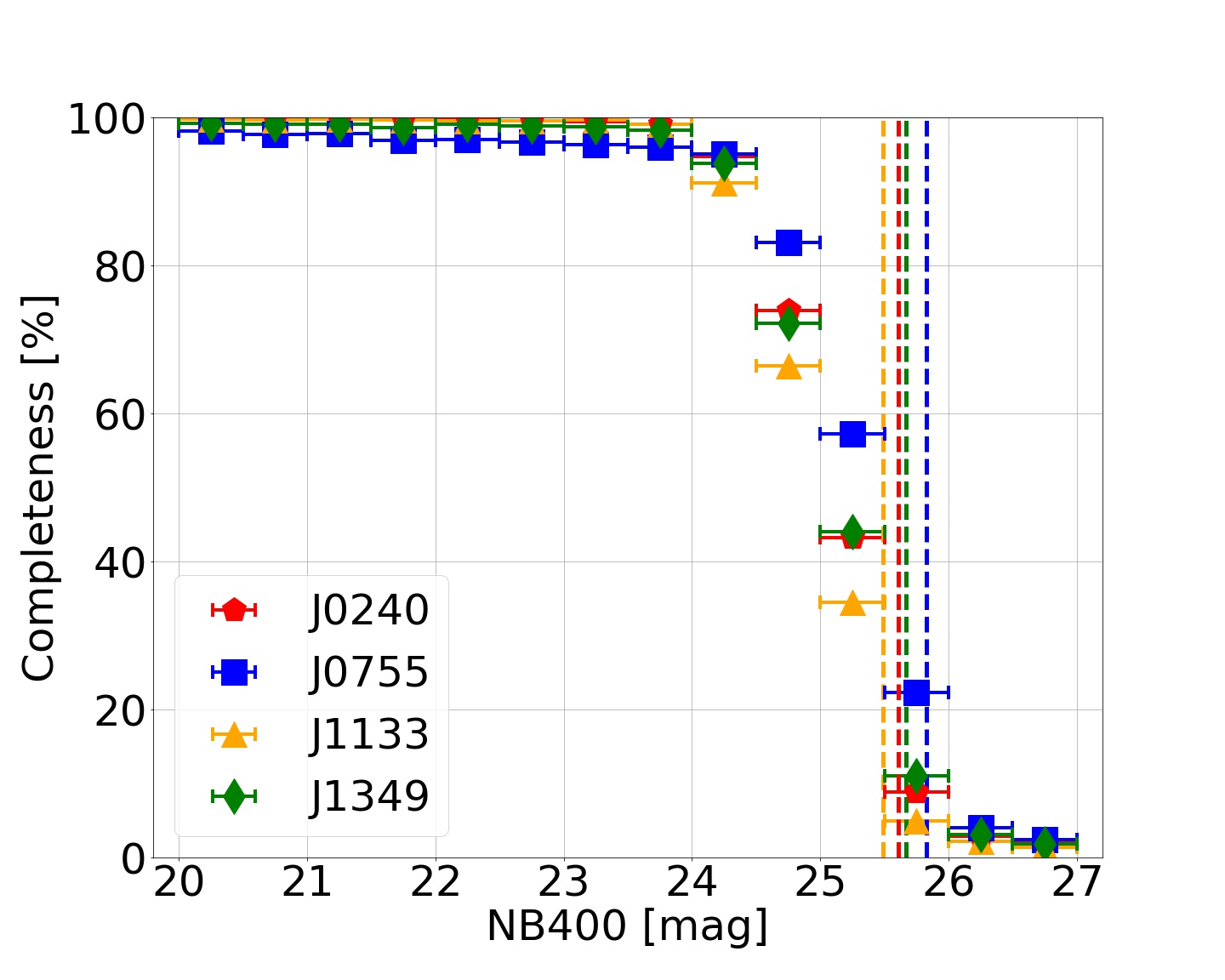}
    \caption{Detection completeness, $f_{det}$, of NB387 (top) and NB400 (bottom) images. The completeness is calculated for each NB387/400 magnitude bin with width $\Delta m = 0.5$ mag. The red pentagon, blue square, yellow triangle, and green diamond represent the detection completeness of J0210, J0222, J0924, and J1419 fields on the top panel, and J0240, J0755, J1133, and J1349 fields in the bottom panel, respectively. Colored dashed lines are the corresponding $5 \sigma$ limiting magnitudes for each field in both panels.}
    \label{fig:fdets}
\end{figure}

The detection completeness $f_{det}$ is estimated by Monte Carlo simulations with the GALSIM package \cite[][]{Rowe_etal_2015}, following the procedure described in previous studies \cite[e.g.][]{Konno_etal_2016, Konno_etal_2018, Itoh_etal_2018}. 
Mock galaxies are constructed according to 14 specified NB387/400 magnitude bins with centers ranging from 20.25 to 26.75 with a bin width $\Delta m = 0.5$. 
Source detection and photometry of the mocks are performed using SExtractor 2.25.0 \cite[][]{Bertin_Arnouts_1996}, in the same manner as our sample selection (See Section 2). 
The detection completeness $f_{\rm{det}}$ is defined as the ratio between the total number of mocks detected and that of mocks created (i.e. 500 for each bin), as a discrete function of the NB387/400 magnitudes. 
We repeat the process 10 times for each field with varying random seeds and obtain the final value by taking the average. 
The results are displayed in figure \ref{fig:fdets}. 
Values of $f_{det}$ are typically above $90\%$ for the bright-end sources and fall rapidly to zero beyond the 
$5\sigma$ limiting magnitudes. 

\subsection{Contamination}

Possible contaminants that can pass our selection criteria are mainly considered to be: 
[OII] $\lambda 3727$ emitters at $z \sim 0.04$, CIII] $\lambda 1909$ emitters at $z \sim 1$, and CIV $\lambda 1549$ emitters at $z \sim 1.5$. 
For the [OII] emitters, the corresponding survey volume 
is about three orders of magnitude smaller than that of LAEs at $z \sim 2.2$, together with the [OII] LFs at $z \sim 0.1$ investigated by \cite{Ciardullo_etal_2013}, we conclude that the amount of low-$z$ [OII] emitters 
is negligible for our sample. 
For CIII] and CIV contaminants, the rest-frame equivalent width would be larger than 30 \AA, much larger than that of star-forming galaxies. 
To investigate the portion of these CIII] and CIV contaminants in our sample, we perform a spectroscopic follow-up for our LAEs.

We present an estimate of the overall contamination rate using spectroscopic observations with Magellan/IMACS on September 29 and 30th, 2022. 
The Magellan/IMACS is set in the multi-slit spectroscopy mode with the f/2 camera, with a field of view of 12 arcsec in radius. 
Slits have a width of 1.2 arcsec and a length of 8.0 arcsec. We
observe one pointing in each of J0210 and J0222 fields, and two pointings in the J0240 field. 
The on-source exposure times are 7500s and 6000s for NB387 and NB400 LAEs, respectively.

The spectra cover wavelength between $\sim$ 3600 and 5700 \AA\ with a spectral resolution of $\approx7$ \AA. 
Since the spectral resolution is not small enough to resolve the doublet of low-$z$ [OII] contaminants, we only use the spectra with $\geq$ two emission lines when estimating the contamination rate. 
After data reduction, we obtain 120 spectra with detected emission lines at the expected wavelength of Ly$\alpha$. 
22 spectra are confirmed LAEs at $z=2.2$ with $\geq$ 2 emission lines, and 2 spectra are foreground objects. 
Therefore, we estimated that contamination rate is about  2/(2+22)=8\%. 
The narrowband magnitudes of the 24 objects with $\geq$ 2 emission lines are 21.3-25.3 mag.

\subsection{Ly$\alpha$ luminosity functions}

\begin{figure*}[ht]
    \includegraphics[width=17cm]{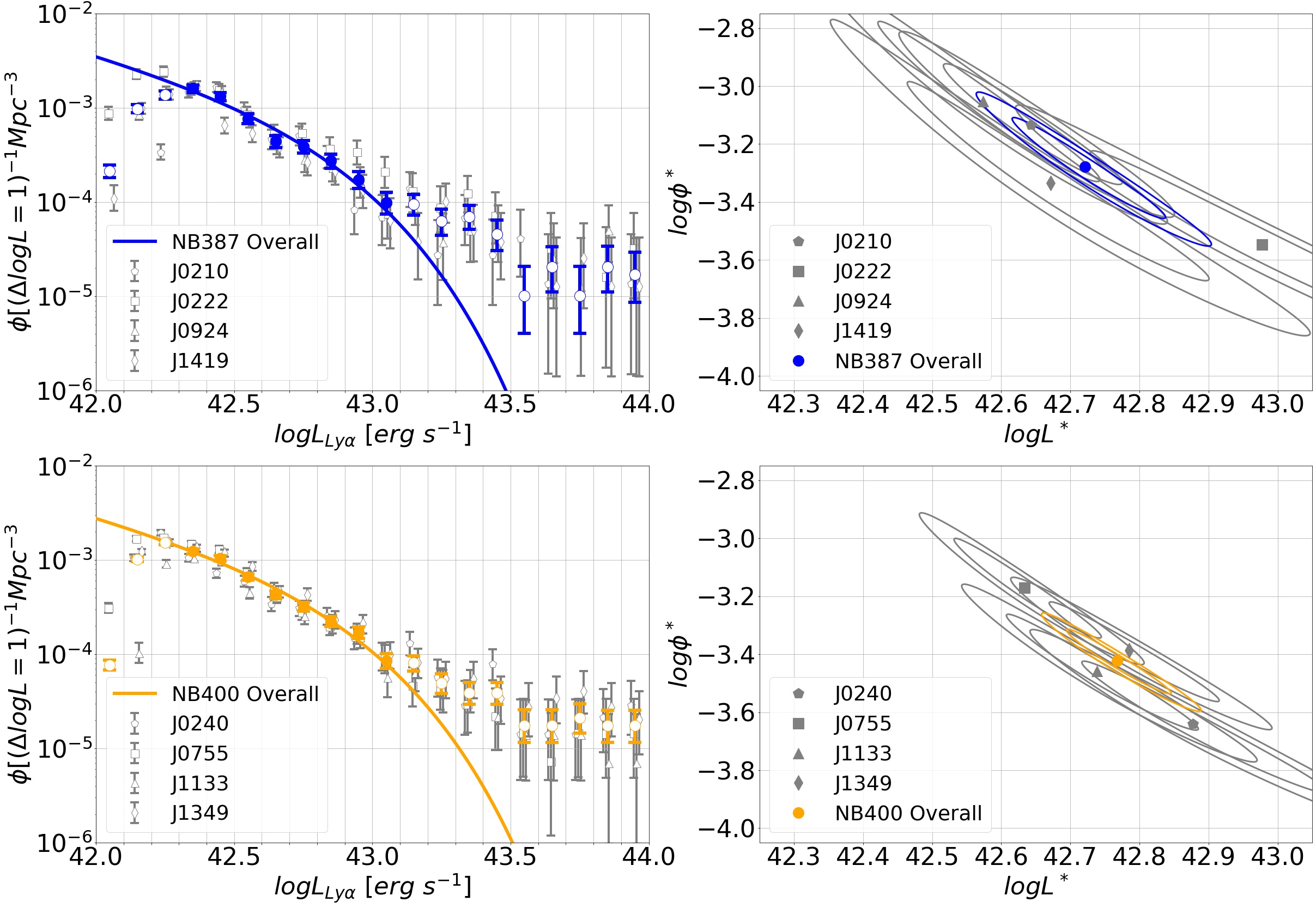}
    \caption{Ly$\alpha$ LFs and error contours for NB387 $\&$ NB400 fields and their corresponding overall results, the points are shifted slightly for clarification. Left panels: LF of each field are shown as the grey data points with different formats specified in the legends, and the overall LFs for 4 NB387 fields (blue) and 4 NB400 fields (yellow) and their corresponding Schechter \cite[][]{Schechter_1976} fits are plotted by solid points and lines. The fitting is done by fixing $\alpha$ at a value concluded by \cite{Konno_etal_2016}, -1.75. During the fitting, we rule out the points with log$L \lesssim 42.3$ which is beyond the completeness correction limit, and the points with log$L \gtrsim 43.1$ which are contaminated by AGNs. These data points are indicated as hollow circles. Right panels: $1\sigma$ and $2\sigma$ error contour of fittings to $L^*$ and $\phi^*$ for 8 fields, NB387 overall and NB400 overall LFs.}
    \label{fig:LFfits_fields}
\end{figure*}

We derive the Ly$\alpha$ LFs in the same manner as many previous studies \cite[e.g.][]{Ouchi_etal_2008, Itoh_etal_2018}. 
We calculate Ly$\alpha$ luminosities using the total magnitude of NB387/400 and g-band, estimated by the AUTO-MAG. 
We calculate the Ly$\alpha$ line flux ($f_{\rm line}$) and the rest-frame UV continuum flux ($f_{\rm c}$) from the NB387/400 and g-band magnitudes ($m_{\rm NB}$ and $m_{BB}$) using the following equation: 
 
\begin{equation}
    m_{\rm NB,BB} + 48.6 = -2.5 {\rm log} \frac{\int_{0}^{\infty} (f_{\rm c} + f_{\rm line})T_{\rm NB,BB} \\ d \nu / \nu}{\int_{0}^{\infty} T_{\rm NB,BB} d\nu / \nu}
    \label{eqn: lineflux}
\end{equation}

Here, $T_{\rm NB}$ and $T_{\rm BB}$ are the transmittance curve in frequency space of the NB and BB filters, respectively. We take the same assumption as \cite{Itoh_etal_2018} that $f_{\rm line}$ is a $\delta$-function and $f_{\rm c}$ is a constant. 

The volume number densities $\phi({\rm logL})$ in a Ly$\alpha$ luminosity bin of $[{\rm logL}-d{\rm logL}/2, {\rm logL}+d{\rm logL}/2]$ is defined as following:
\begin{equation}
    \phi({\rm log}L) d{\rm log}L \\ = \\ \sum_{i} \\ \frac{1}{f_{\rm i, det} \\ V_{\rm eff}}
    \label{eqn: phiLlya}
\end{equation}
where the sum is taken over all LAEs in the specified bin. 
$f_{\rm i,det}$ is the detection completeness for the corresponding NB387/NB400 magnitude of the $i$th LAE, and $V_{\rm eff}$ is the effective survey volume assuming a top-hat filter transmittance. 
We set the bin width to be $d{\rm logL} = 0.1 {\rm dex}$, which is the same as that of \cite{Konno_etal_2016}.
Note that the actual NB387 and NB400 filter transmission curves are not perfect top-hat. \cite{Sobral_etal_2017} investigated the incompleteness due to this effect and found the correction factor range from 0.97 in the faintest bin to 1.3 in the brightest bin. 
This study mainly focuses on the faint end and excludes bright end as described below, therefore, we take the assumption that the filter transmittance shapes of NB387 and NB400 filters are top-hat.\par

The left two panels of figure \ref{fig:LFfits_fields} show LFs derived from 8 fields, as well as the overall LFs of NB387 and NB400. 
The error bars include the Poisson uncertainties calculated from \cite{Gehrels_1986}, as well as the field-to-field variance estimated from our measurement in section 4. 
The values are 0.063 for NB387 and 0.034 for NB400 (See section 4.2 for details). 
We include the field-to-field variance in this section.
The low luminosity point has been excluded for low completeness. 
Note that we choose the value of log$L > 42.3$ for all fields for keeping the fitting process consistent.
Using the minimum $\chi^2$ fitting, the LFs are fitted to a Schechter function \cite[][]{Schechter_1976}, defined as:

\begin{equation}
    \phi({\rm log}L) = {\rm log}_e(10) \\ \phi^* \\ (\frac{L}{L^*})^{\alpha+1} \\ e^{-\frac{L}{L^*}}
    \label{eqn: Schechter}
\end{equation}

Here, Schechter parameters $\alpha$, $L^*$, and $\phi^*$ characterize the 
faint-end slope, luminosity, and density respectively \cite[][]{Schechter_1976}. \cite{Konno_etal_2016} surveyed the Lya LFs for z$\sim$2.2 LAEs using the NB387 data from Subaru/Suprime Cam \cite[][]{Miyazaki_etal_2014} in five fields and fitted three Schechter parameters simultaneously as free parameters. 
They reported a faint-end slope $\alpha$ of $-1.75_{-0.09}^{+0.10}$, since this value is reasonably well-constrained \cite[][]{Konno_etal_2016}, we fix our $\alpha$ at this value. 
The goal of this study is to investigate the field-to-field variation, and thus having $\alpha$ fixed at the same value has the benefit of reducing differences in the expected galaxy abundance for all fittings, so that the statistical fluctuations become significant. 
The right panels of figure \ref{fig:LFfits_fields} show the $68\%$ and $90\%$ confidence levels, as well as the best-fit values of the other two parameters $L^*$ and $\phi^*$. 
The best-fitting results are demonstrated in table \ref{tab:LFfits}.

\begin{table}[h!]
\centering
\begin{tabular}{c c c} 
 \hline
  & L$^*$ [10$^{42}$ erg s$^{-1}$] & $\phi^*$ [10$^{-4}$ Mpc$^{-3}$] \\ [0.5ex] 
 \hline\hline
 J0210 & $4.39_{-1.11}^{+1.56}$ & $7.37_{-2.81}^{+4.59}$  \\ [1ex] 
 J0222 & $9.49_{-4.15}^{+14.99}$ & $2.84_{-1.81}^{+3.15}$ \\[1ex] 
 J0924 & $3.75_{-1.11}^{+1.79}$ & $8.86_{-4.19}^{+7.88}$ \\[1ex] 
 J1419 & $4.69_{-1.78}^{+3.30}$ & $4.63_{-2.50}^{+5.79}$  \\[1ex] 
 J0240 & $7.53_{-2.33}^{+4.36}$ & $2.28_{-0.97}^{+1.49}$  \\[1ex] 
 J0755 & $4.30_{-0.90}^{+1.25}$ & $6.75_{-2.23}^{+3.24}$  \\[1ex] 
 J1133 & $5.48_{-1.44}^{+2.21}$ & $3.48_{-1.30}^{+2.01}$  \\[1ex] 
 J1349 & $6.10_{-1.43}^{+2.14}$ & $4.10_{-1.37}^{+1.94}$  \\
 \hline
 NB387 & $5.26_{-1.13}^{+1.63}$ & $5.26_{-1.77}^{+2.52}$  \\[1ex] 
 NB400 & $5.85_{-0.91}^{+1.15}$ & $3.78_{-0.87}^{+1.09}$  \\ [1ex] 
 Overall & $5.18_{-0.40}^{+0.43}$ & $4.87_{-0.55}^{+0.54}$ \\
 \hline
\end{tabular}
\caption{The best-fit Schechter parameters for the Ly$\alpha$ LFs of individual fields and the overall data of NB387, NB400, and all 8 fields. All fittings are performed within a log$L$ range of 42.3-43.1 erg s$^{-1}$ and the faint-end slopes are fixed at -1.75.}
\label{tab:LFfits}
\end{table}

We combine NB387 and NB400 data to compute the overall LF for all 8 fields (See red points and line in figure \ref{fig:LFlits}). 
Due to the redshift difference between NB387 and NB400, we argue that the difference in the luminosity function between NB387 and NB400 is due to field-to-field variation rather than redshift evolution. 

An excess in number densities at the bright-end with log$L \gtrsim 43.1$ is presented in the left panels of figure \ref{fig:LFfits_fields}. 
One possible explanation is the magnification bias \cite[e.g.][]{Mason_etal_2015}, the luminosity of several LAEs is raised due to gravitational lensing by foreground galaxies. 
However, past studies  \cite[e.g.][]{Konno_etal_2016,Sobral_etal_2017} suggest that the number density excess at the bright end is due to type-I AGN rather than magnification bias. 
\cite{Zhang_etal_2021} have also confirmed this conclusion using the HETDEX spectroscopic surveys. 
\cite{Konno_etal_2016} stated that, due to the presence of significantly smaller error bars at the faint end, the humps at bright end do not affect the fitting results significantly. 
We attempted to fit a Schechter function to our overall LF both with ($42.3<$logL$<44.1$) and without the bright-end (42.3$<$logL$<$43.1), and obtained two sets of best-fit parameters of ($\alpha=-1.75$ (fixed), $L^{*}=6.23_{-0.54}^{+0.61}$ $ \times 10^{42}$ erg s$^{-1}$, $\phi^{*}=3.81_{-0.43}^{+0.51}$ $ \times 10^{-4}$ Mpc$^{-3}$) and ($\alpha=-1.75$ (fixed), $L^{*}=5.18_{-0.40}^{+0.43}$ $ \times 10^{42}$ erg s$^{-1}$, $\phi^{*}=4.87_{-0.55}^{+0.54}$ $ \times 10^{-4}$ Mpc$^{-3}$).
We thus conclude that the bright-end hump does not affect our fitting significantly.
We chose to perform curve-fitting without the bright-end hump to make the following cosmic variance investigation as accurate as possible. 
This results in a log$L_{{\rm Ly}\alpha}$ coverage range of $42.3-43.1$. 

Further, we also fit a power law $log_{10} \phi = A \times log_{10}(L) + B$ according to \cite{Sobral_etal_2017} to the bright-end LF with log$L > 43.1$, and obtain best-fit parameters $A = -1.16_{-0.22}^{+0.15}$ and $B = 46.04_{-6.31}^{+9.73}$. We show this power-law fitting using the red dashed line in figure \ref{fig:LFlits}.

\subsection{Comparison with previous studies} 

\begin{figure}[ht]
    \includegraphics[width=8.75cm]{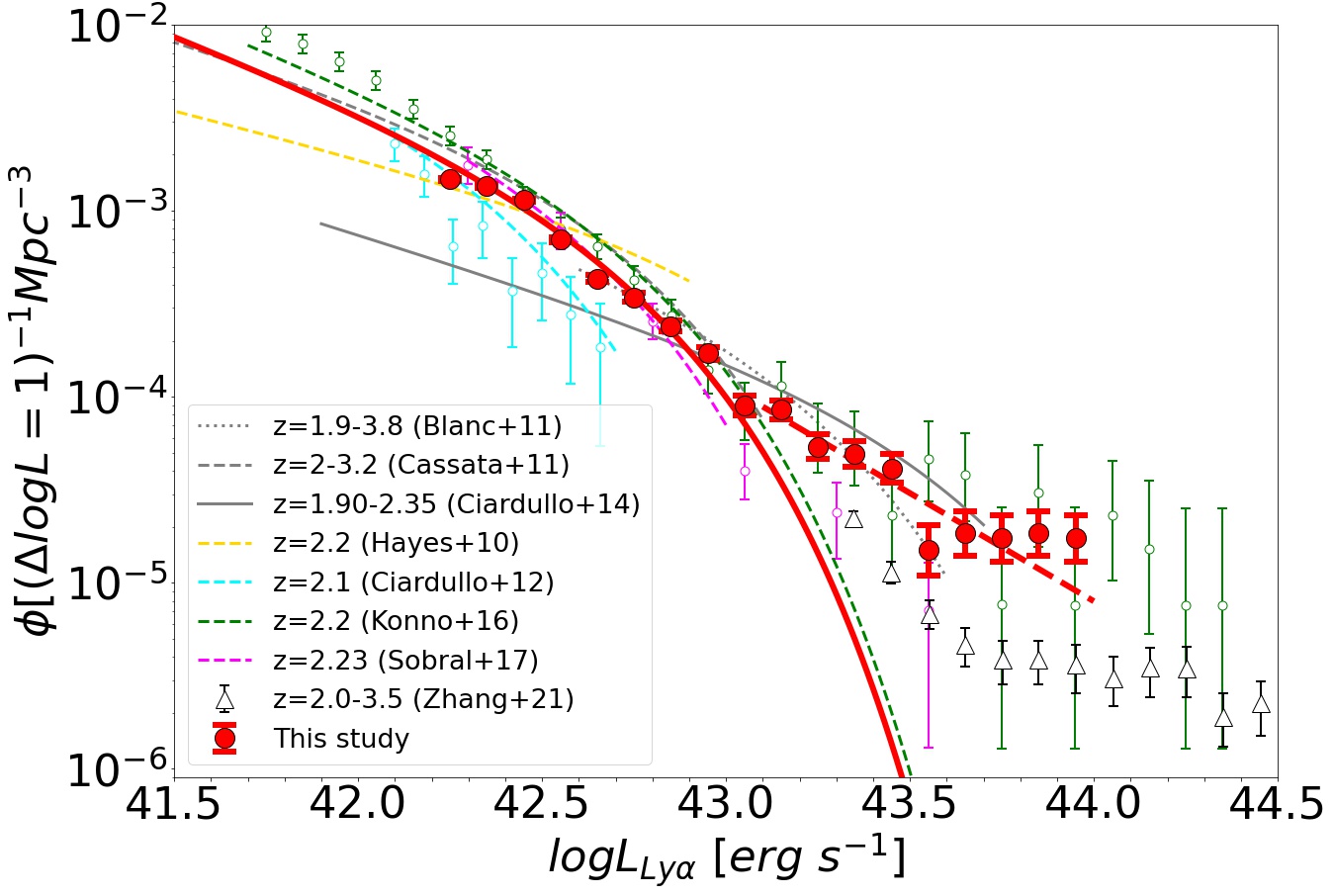}
    \caption{Comparison of our Ly$\alpha$ LFs with previous studies of LAEs at $z \sim 2$. The red filled circles, red solid line, and red dashed line are the LF data points, the best-fit Schechter function, and the best-fit power law function derived from our overall sample respectively. The black dotted, dashed and solid lines represent the best-fit Schechter functions obtained by spectroscopic surveys of \cite{Blanc_etal_2011,Cassata_etal_2011} and \cite{Ciardullo_etal_2014}, at redshift range $1.9<z<3.8$, $2.0<z<3.2$, and $1.90<z<2.35$ respectively. The gold, cyan, green, and magenta hollow circles and dashed lines are the data points and the best-fit Schechter functions from narrowband imaging surveys by \cite{Hayes_etal_2010, Ciardullo_etal_2012,Konno_etal_2016} and \cite{Sobral_etal_2017} respectively. We show the lines within the luminosity ranges limited by corresponding studies. The black hollow triangles are LFs computed by \cite{Zhang_etal_2021}, which confirmed that the bright-end hump is mainly caused by AGN.}
    \label{fig:LFlits}
\end{figure}

We show the comparison between our Ly$\alpha$ LFs and that of previous studies in figure \ref{fig:LFlits}. 
Results from blank-field spectroscopic surveys \cite[][]{Blanc_etal_2011,Cassata_etal_2011,Ciardullo_etal_2014} are plotted with black lines with different line styles, while the results from narrowband imaging surveys \cite[][]{Hayes_etal_2010,Ciardullo_etal_2012,Konno_etal_2016,Sobral_etal_2017} are plotted with dashed lines with different colors. 
We also plot Ly$\alpha$ LF from the HETDEX spectroscopic survey by \cite{Zhang_etal_2021} using black triangles. 
They have confirmed that type I AGNs can fully explain the bright-end hump. 

\begin{table*}[t]
\centering
\begin{tabular}{c c c c c c} 
 \hline
  & $\alpha$ & L$^*$ [10$^{42}$ erg s$^{-1}$] & $\phi^*$ [10$^{-4}$ Mpc$^{-3}$] & A$_{tot}$ [deg$^2$] & V$_{tot}$ [10$^6$ Mpc$^3$]\\ [0.5ex] 
 \hline\hline
  This study & -1.75 (fixed) & $5.18_{-0.40}^{+0.43}$ & $4.87_{-0.55}^{+0.54}$ & 11.62 & 8.71 \\ [1ex] 
  \cite{Blanc_etal_2011} & -1.7 (fixed) & $16.3_{-10.8}^{+94.6}$ & $1.0_{-0.9}^{+5.4}$ & 0.047 & 1 \\[1ex] 
  \cite{Cassata_etal_2011} & -$1.6_{-0.12}^{+0.12}$ & 5.0 (fixed) & $7.1_{-1.8}^{+2.4}$ & 0.78 & 2.3 \\[1ex]
  \cite{Ciardullo_etal_2014} & -1.6 (fixed) & $39.8_{-16.4}^{+98.2}$ & 0.36 & 0.044 & 1.03 \\[1ex]
  \cite{Hayes_etal_2010} & -$1.49_{-0.27}^{+0.27}$ & $14.5_{-7.54}^{+15.7}$ & $2.34_{-1.64}^{+5.42}$ & 0.016 & 0.005 \\[1ex]
  \cite{Ciardullo_etal_2012} & -1.65 (fixed) & $2.14_{-0.52}^{+0.68}$ & $13.8_{-1.5}^{+1.7}$ & 0.36 & 0.17 \\[1ex]
  \cite{Konno_etal_2016} & -$1.75_{-0.09}^{+0.10}$ & $5.29_{-1.13}^{+1.67}$ & $6.32_{-2.31}^{+3.08}$ & 1.43 & 1.32 \\[1ex]
  \cite{Sobral_etal_2017} & -$1.75_{-0.25}^{+0.25}$ & $3.89_{-0.65}^{+1.73}$ & $9.13_{-4.41}^{+3.09}$ & 1.43 & 0.73 \\[1ex]
 \hline
\end{tabular}
\caption{The best-fit Schechter parameters for the Ly$\alpha$ LFs of previous z$\sim$2.2 LAEs, and their corresponding total survey area and survey volumes.}
\label{tab:LFlits}
\end{table*}

The LAE samples used by \cite{Blanc_etal_2011} and \cite{Ciardullo_etal_2014} are obtained from spectroscopic observations from the Hobby Eberly Telescope Dark Energy Experiment (HETDEX) Pilot Survey, while \cite{Cassata_etal_2011}'s sample comes from the VIMOS survey \cite{LeFevre_etal_2003}. 
The redshift range covered by these spectroscopic surveys is $1.9<z<3.8$, $2.0<z<3.2$, and $1.90<z<2.35$ respectively, with a Ly$\alpha$ EW$_0$  greater than 20 Å for most LAEs. 
\cite{Hayes_etal_2010} investigated Ly$\alpha$ LFs as well as the escape fraction of z=2.2 LAEs in the GOODS-South field, obtained by narrowband imaging using NB388 filter for VLT/FORS1. 
Ly$\alpha$ LFs from \cite{Ciardullo_etal_2012} are derived from z=2.1 LAE samples of \cite{Guaita_etal_2010}, which is obtained from an ultra-deep narrowband MUSYC in the ECDF-S field, with central wavelength 3727 Å. 
We also over-plot the data points from \cite{Zhang_etal_2021}, who confirmed the bright-end hump of z$\sim$2 LAE LFs are mainly caused by type I AGNs. 

\cite{Konno_etal_2016} investigated Ly$\alpha$ LFs of $z \sim 2.2$ LAEs using the NB387 data of Subaru/Suprime Cam (SC) and constructed their full LAE sample from the SXDS, COSMOS, CDFS, HDFN, and SSA22 fields. 
\cite{Sobral_etal_2017} carried out deep narrowband imaging with a custom-built NB392 filter for the Isaac Newton Telescope’s Wide Field Camera (INT/WFC), they derived Ly$\alpha$ LFs from $z\sim2.23$ LAEs detected in the COSMOS and UDS fields. 
The Ly$\alpha$ EW$_0$ criteria of color excess in narrowband in these narrowband imaging surveys are all EW$_0$=20 Å, which are consistent with our selections. 
The best-fit Schechter parameters of the literature mentioned above and that of our studies are summarised in table \ref{tab:LFlits}. 

In figure \ref{fig:LFlits}, our results show a small offset with \cite{Konno_etal_2016}.
Apart from the result of \cite{Hayes_etal_2010} which subjects to a huge uncertainty owing to small statistics, all Ly$\alpha$ LFs from the narrowband imaging surveys yield similar of $\alpha$ and $L^*$, but small deviations in $\phi^*$. 

\section{COSMIC VARIANCE}
\subsection{Measuring the field-to-field variations}

In previous studies \cite[e.g.][]{Konno_etal_2016,Ouchi_etal_2008}, the field-to-field variation is characterized by $\sigma_g$, which represents the expected dispersion in galaxy counts owing to the cosmic variance. 
The expected galaxy counts $N$ can be obtained by integrating LFs over the effective survey volume probed by a field \cite[e.g.][]{Robertson_2010a}.
Therefore, we make a series of this value $N$ by assigning pointing on each field image, with the coverage smaller than one  HSC field of view. 
For each pointing assigned, a Ly$\alpha$ LF can be derived from a subsample consisting of LAEs within that pointing. 
By integrating the best-fit Schechter functions, the galaxy abundance $N$ of each pointing can be obtained: 

\begin{equation}
    N/V_{\rm eff} = \int_{{\rm log}L_{\rm lim}}^{\infty} \phi({\rm log}L) \\ d{\rm log}L
    \label{eqn:IntSchechter}
\end{equation}
Here, log$L_{\rm lim}$ corresponds to the integration limit for Ly$\alpha$ luminosities. We adopt the value from \cite{Konno_etal_2016} $logL_{\rm lim} = 41.41$ erg s$^{-1}$ for all fittings. 
Moreover, $logL_{\rm lim}$ has also been discussed in Ly$\alpha$ luminosity densities \cite[e.g.][]{Konno_etal_2016,Ouchi_etal_2008}, which characterize the mean Ly$\alpha$ luminosity per unit volume of the pointing. 
Different choices of log$L_{\rm lim}$ can result in systematic errors in the integral, thus shifting the position of mean galaxy abundances, but leaving the distribution unaffected. 

To keep the calculation process of galaxy abundances $N$ consistent for each pointing, we apply the same fitting procedure for each pointing as follows: 
fixing the faint-end slope $\alpha$ at -1.75, as mentioned in section 3.3, and excluding the data points with log$L < 42.3$, which are beyond the completeness limit, and the data points with log$L > 43.1$, contaminated by AGNs. 
The lower limit for integrating LFs is set to be $logL_{\rm lim} = 41.41$. 
In addition, since we aim to investigate the field-to-field variation itself, we only include Poisson uncertainties in the error bars when fitting the LFs for this section.

Furthermore, MAMMOTH fields are selected by targeting overdensities. 
Overdensities could bias the field-to-field variations which are supposed to measure in random fields. 
To reduce the bias on the cosmic variance measurements caused by these overdensities, we clip out the regions pre-selected by MAMMOTH on each field image. 
Among our 8 fields, six (J0210, J0222, J0924, J1419, J1133, and J1349) are selected targeting coherently strong Ly$\alpha$ absorption systems (CoSLAs) \cite[See][]{Liang_etal_2021, Cai_etal_2016}, two (J0240 and J0755) are selected targeting at grouping quasars (See Cai et al. in prep.), we refer these as the MAMMOTH targets.
The MAMMOTH pre-selected regions are defined to cover these MAMMOTH targets, with an aperture size of 15 comoving Mpc, which is the typical scale of protoclusters at z $\sim$ 2.2 \cite[see e.g., ][]{Chiang_etal_2013}. 
These regions are treated as masks and the LAEs within these regions are excluded when sampling the pointings.

The pointings are assigned with circular shapes and fixed volume coverages. 
We scale up the pointing areas assigned in the NB387 field images by a factor of depth$_{\rm NB400}$/depth$_{\rm NB387}$, so that the volumes probed by pointings in NB387 images is the same as those in NB400 images. 
The centers are created uniformly within the field of view for each field. 
We calculate the galaxy number density $n_i$ for the $i$-th pointing as follows:

\begin{equation}
    n_i = \int_{logL_{\mbox{\tiny lim}}}^{\infty} \phi_i(logL) \\ dlogL
    \label{eqn:gad}
\end{equation}

where $\phi_i(logL)$ is the best-fit Schechter function for the $i$-th pointing. 
In such a way, the set $\{n_i\}$ can be considered to have the same effective survey volume $\overline{V}_{\rm eff}$. 
There are two reasons we quantify galaxy numbers by integrating the LFs instead of directly using the cumulative numbers. 
One is to investigate the effect of cosmic variance on LF itself, the other is that this approach can take the completeness correction into account.

Log-normal distributions provide an excellent fit for cosmological density distribution function from CDM non-linear dynamics \cite[][]{Bernardeau_Kofman_1995}. 
The skewness behaviour towards zero can effectively imitate the dark matter fluctuations at the high-mass end \cite[][]{Coles_Jones_1991}. 
\cite{Yang_etal_2010} investigated Ly$\alpha$ nebulae in CDFS, CDFN, and COSMOS fields, and reported a strong field-to-field variance. 
They quantified the field-to-field variance by assuming a log-normal distribution for their Ly$\alpha$ blob counts. 
Here we adopted a similar model, which assumes the galaxy number density $n$ follows a log-normal distribution $n \sim \mbox{Log-N}(\overline{n},\sigma_{LN}^2)$, where $\sigma_{LN}^2$ is the log-normal variance 
and is related to the actual variance by $\sigma_v^2 = exp(\sigma_{LN}^2)-1$. 
Since the dispersion of $N$ can be estimated as the field-to-field variance plus the Poisson variance (i.e. $\sigma_g^2 \overline{N}^2 + \overline{N}$, see e.g. \cite{Robertson_2010a}), we then estimate $\sigma_g$ using the relation:
\begin{equation}
    \sigma_g^2 = {\rm{exp}}(\sigma_{LN}^2)-1-\frac{1}{\overline{n} \overline{V}_{\mbox{\tiny eff}}}
    \label{eqn:sigg}
\end{equation}
where the last term corresponds to the Poisson variance. 
$\overline{n}$ and $\overline{V}_{\mbox{\tiny eff}}$ correspond to the mean values of galaxy number density and effective survey volume respectively.

We fit each set of $\{ n_i\}$ to this log-normal distribution:
\begin{equation}
    \phi({\rm log}n) = \frac{C}{\sqrt{2 \pi \sigma_{LN}^2}} {\rm{exp}}(- \frac{({\rm log}n - {\rm log}\overline{n})^2}{2 \sigma_{LN}^2})
    \label{eqn:LN1}
\end{equation}
where $\phi({\rm log}n)$ is defined as the number counts of log$n$ within a specified bin divided by the total number, per bin width. 
We also take the error ranges of $n$ for each pointings into account, and obtain $\phi({\rm log}n)$ weighted by the reciprocal of these error ranges. 
$C$ is an arbitrary normalization factor. 
We fit the three parameters $C$, log$\overline{n}$ and $\sigma_{LN}^2$ in equation \ref{eqn:LN1} simultaneously, and obtain a best-fit Gaussian mean log$\overline{n}_{\rm field}$ for each field. 

To eliminate the diversity in the Gaussian mean between different fields, we take the best-fit $\overline{n}_{\rm field}$ from fittings for individual fields, and compute a quantity log$(\frac{n}{\bar{n}})_{\rm field}$ for each field. 
This quantity should have the same mean value of 0 for all fields, we thus fit the following equation to the overall distributions of log$(\frac{n}{\bar{n}})$ combining multiple fields:
\begin{equation}
    \phi({\rm log}\frac{n}{\bar{n}}) = \frac{C}{\sqrt{2 \pi \sigma_{LN}^2}}\ {\rm{exp}}(- \frac{({\rm log}\frac{n}{\bar{n}})^2}{2 \sigma_{LN}^2})
    \label{eqn:LN2}
\end{equation}
where $\sigma_{LN}^2$ and the arbitrary normalizing factor $C$ are treated as free parameters for fitting. 
The lower and upper limit for the errors of log$(\frac{n}{\bar{n}})$ is calculated as log$\frac{n_{\rm lower}}{\overline{n}_{\rm upper}}$ and log$\frac{n_{\rm upper}}{\overline{n}_{\rm lower}}$ respectively. 
Furthermore, we scale back each $\phi($log$\frac{n}{\bar{n}})_{\rm field}$ according to the normalizing factor $C_{\rm field}$ obtained in equation \ref{eqn:LN1} when combining $ \{ {\rm log}(\frac{n}{\bar{n}})_i \}_{\rm field}$ for different fields.

\subsection{Volume dependence of $\sigma_g$}

In the sections above, we described our approach to obtain $\sigma_g$ by randomly assigning pointings with fixed volume coverages for 8 fields, we investigate the effects on $\sigma_g$ by changing the effective volume of pointings in this section. \par

To obtain the relation between $\sigma_g$ and $V_{\rm eff}$, we need to keep $V_{\rm eff}$ similar for every pointing during each investigation.
In the sections above, we randomly create our pointing centers across the whole circular HSC field of view and clip out the MAMMOTH-specified regions. 
However, this results in a relatively large scatter in the effective survey volumes $V_{\rm eff}$, since pointings at the edge lack full coverage of the field image and pointings overlapping with masked and clipped regions have smaller $V_{\rm eff}$ than expected. 
To deal with this effect, we first generate pointing centers within a confined circle with a radius defined as, the radius of HSC field of view minus the aperture size set for the pointings. 
In this way, we ensure that every pointings fully cover the field image. 
Next, we repeat the process described in section 4.1 with various sizes of pointings, and obtain multiple pairs of ($V_{\rm eff}$, log$(\frac{n}{\bar{n}})$), with each pair corresponding to a pointing assigned. 
We then make bins according to the $V_{\rm eff}$ distribution. 
For each $V_{\rm eff}$ bin, we calculate the mean value of the effective volumes $\overline{V}_{\rm eff}$ and $\sigma_g$ by fitting the Gaussian to log$(\frac{n}{\bar{n}})$ of the pointings whose $V_{\rm eff}$ lies inside the bin.

\begin{figure}[ht]
    \centering
    \includegraphics[width=8cm]{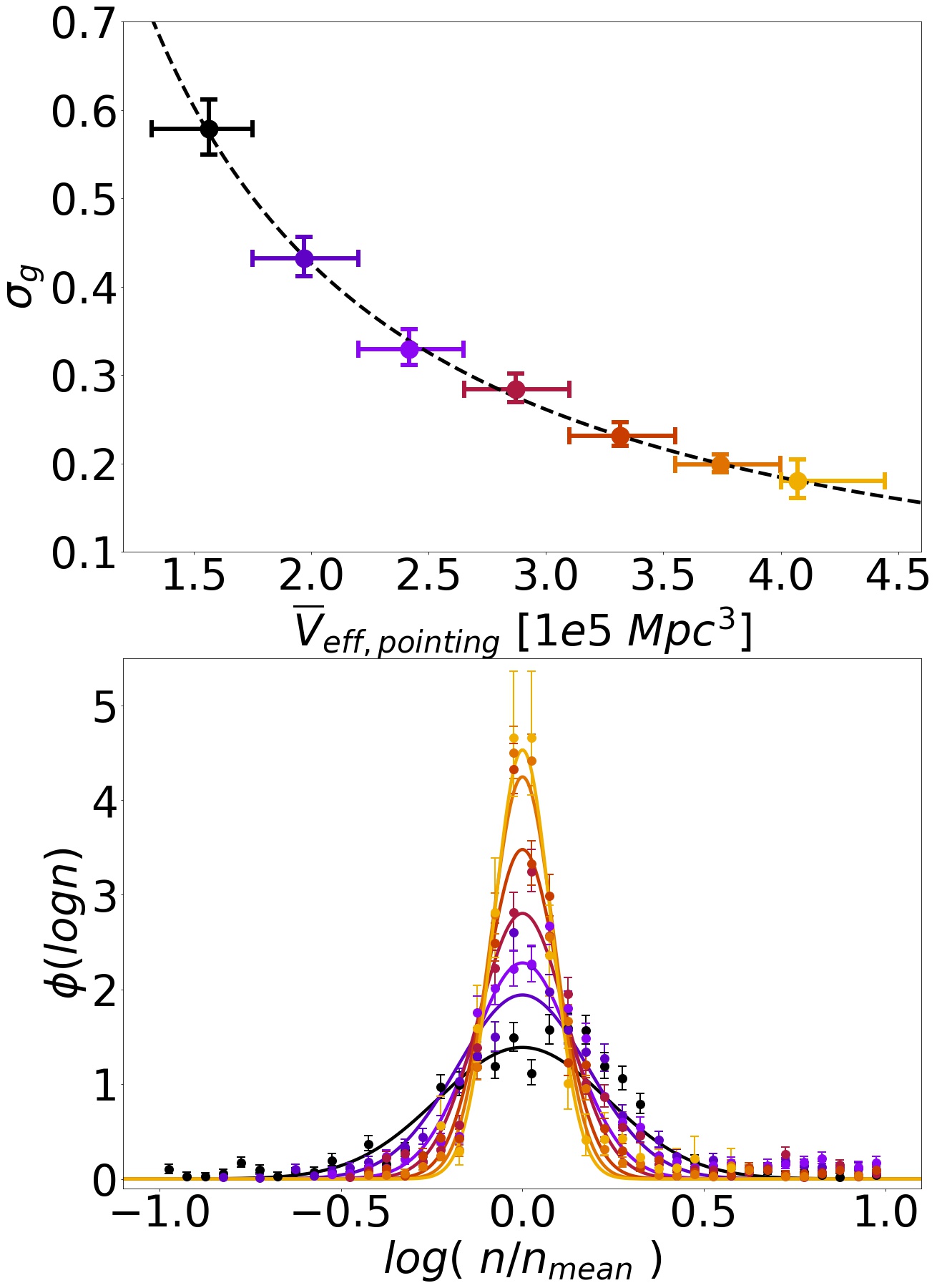}
    \caption{Log-normal fits of the overall $\{ (\frac{n}{\bar{n}})_i\}$ distribution for 8 fields, with varying mean effective volume per bin $\overline{V}_{\rm eff}$. The bottom panel shows the distributions of $\{ {\log}(\frac{n}{\bar{n}})_i\}$, and the corresponding best-fit Gaussian function (equation \ref{eqn:LN2}). Each $V_{\rm eff}$ bin is indicated by different colors shown in the color bar. The top panel shows $\sigma_g$ obtained by each investigation plotted against $\overline{V}_{\rm eff}$, we fit a simple power-law $\sigma_g = k \times (\frac{V_{\rm eff}}{10^5 {\rm Mpc}^3})^{\beta}$, and demonstrate it using a black dashed line. The best-fit $\beta$ and k has a value of $-1.209_{-0.106}^{+0.106}$ and $0.986_{-0.100}^{+0.108}$ respectively. The vertical and horizontal error bars show the 1-$\sigma$ error ranges of $\sigma_g$ adopted from the Gaussian fitting and the coverages of each $V_{\rm eff}$ bin respectively. Only vertical error bars are taken into account when performing the power-law fitting.}
    \label{fig:CV_varyV}
\end{figure}

We plot $\sigma_g$ against $\overline{V}_{\rm eff}$ and show the results in figure \ref{fig:CV_varyV}. 
The range of $\overline{V}_{\rm eff}$ is limited by the survey depths and effective survey areas of the field images. 
A too small $\overline{V}_{\rm eff}$ will cause the pointings to pick up a small number of LAEs or even zero LAEs, which leads to a poor Schechter fit of LFs. 
On the other hand, a large $\overline{V}_{\rm eff}$ value that covers a whole field image will create a delta function for its distribution, which is meaningless for $\sigma_g$ investigation. 
The smallest size assigned for the pointings is 0.2 deg$^2$ for NB400 and 0.34 deg$^2$ for NB387 (The depths of the two filters are different and the areas are changed accordingly to keep the volume constant), the corresponding mean number of LAEs per pointing is $\sim$70. 
The largest size assigned for the pointings is 0.5 deg$^2$ for NB400 and 0.84 deg$^2$ for NB387, the corresponding mean number of LAEs per pointing is $\sim$170.
This results in a volume coverage between $1.86 \times 10^5 Mpc^{-3}$ and $4.61 \times 10^5 Mpc^{-3}$ for the pointings (Although, due to the overlapping between pointings and masked or clipped regions, the actual values of $\overline{V}_{\rm eff}$ are smaller than expected in general).
Detailed information on each $V_{\rm eff}$ bins is shown in table \ref{tab:LNvaryV}.

\begin{table}[h!]
\centering
\begin{tabular}{c c c c} 
 \hline
 $\overline{V}_{\rm eff}$ [10$^5$ Mpc$^3$] & N$_{pointings}$ & $\sigma_{LN}^2$ & $\sigma_g$ \\ [0.5ex] 
 \hline\hline
 1.57 & 1133 & $0.290_{-0.025}^{+0.027}$ & $0.578_{-0.029}^{+0.033}$ \\[1ex]
 1.97 & 1313 & $0.172_{-0.015}^{+0.016}$ & $0.432_{-0.021}^{+0.024}$ \\[1ex]
 2.42 & 1302 & $0.104_{-0.010}^{+0.012}$ & $0.329_{-0.018}^{+0.023}$ \\[1ex]
 2.87 & 1324 & $0.078_{-0.007}^{+0.008}$ & $0.284_{-0.014}^{+0.018}$ \\[1ex]
 3.32 & 1201 & $0.053_{-0.005}^{+0.006}$ & $0.231_{-0.012}^{+0.016}$ \\[1ex]
 3.74 & 1133 & $0.039_{-0.003}^{+0.003}$ & $0.199_{-0.009}^{+0.012}$ \\[1ex]
 4.07 & 209  & $0.033_{-0.006}^{+0.008}$ & $0.180_{-0.019}^{+0.024}$ \\[1ex]

 \hline
\end{tabular}
\caption{The information of each $V_{\rm eff}$ bin. From left to right, the columns are: mean value of $V_{\rm eff}$ inside the bin, number of pointings within the $V_{\rm eff}$ bin, best-fit $\sigma_{LN}^2$ of the log-normal fitting, and the resulting $\sigma_g$, respectively. Note that number of pointings in the last bin is relatively smaller than those in other bins, this results in a larger error of the $\sigma_g$ calculated, which can also be seen in figure \ref{fig:CV_varyV}. }
\label{tab:LNvaryV}
\end{table}

In previous studies \cite[e.g.][]{Ouchi_etal_2008,Konno_etal_2016,Robertson_2010a}, $\sigma_g$ is also defined by the following:
\begin{equation}
    \sigma_g = b_g \sigma_{DM}(R,z)
    \label{eqn:CV_bias}
\end{equation}
where $b_g$ is the bias parameter of galaxies, defined as the ratio between the galaxy and matter correlation functions $\xi_{gg}/\xi_{mm}$ \cite[][]{Robertson_2010a}, and $\sigma_{DM}^2(R,z)$ is the dark matter fluctuation in a sphere with radius $R$ at redshift z. 
This suggests that $\sigma_g$ only depends on the survey volume at a given redshift at large scales. 
$\sigma_{DM}$ can be calculated as following \cite[e.g.][]{Newman_Davis_2002,Moster_etal_2010}:
\begin{equation}
    \sigma_{DM}^2(V,z) = \frac{1}{V^2} \int_{0}^{R} \xi (|\vec{r_1}-\vec{r_2}|) d V_1 d V_2
    \label{eqn:sigDM_R}
\end{equation}
where $\xi$ is the two-point correlation function of galaxies, which can be treated as a power-law $\xi(r) = (r_0/r)^{\gamma}$ in previous studies \cite[e.g.][]{Somerville_etal_2004,Moster_etal_2010}. 
This makes equation \ref{eqn:sigDM_R} analytically solvable into a closed form: $\sigma_{DM}^2 = C(r_0/r)^{\gamma}$ \cite[][]{Somerville_etal_2004} Since the bias parameter $b_g$ only depends on the redshift, the volume-dependency of $\sigma_g$ comes from $\sigma_{DM}$ only. 
In this study, we assume a simple power-law $\sigma_{DM}^2 \propto V^{\beta}$, and thus fit our $\sigma_g$ and $\overline{V}_{\rm eff,pointing}$ to a simple relation $\sigma_g = k \times (\frac{V_{\rm eff}}{10^5 {\rm Mpc}^3})^{\beta}$ with some constant $\beta$ and k. 
We find a best-fit value of $-1.209_{-0.106}^{+0.106}$ for $\beta$ and $0.986_{-0.100}^{+0.108}$ for k. 
This is shown as the black dashed line in figure \ref{fig:CV_varyV}. 
Using this relation, we find a $\sigma_g$ of $0.063_{-0.018}^{+0.026}$ for NB387 with survey volume $9.72 \times 10^5$ Mpc$^3$, and $0.034_{-0.006}^{+0.016}$ for NB400 with survey volume $16.34 \times 10^5$ Mpc$^3$. 
We use these two values as the field-to-field variance when reporting LF results in section 3 (See section 3.3).

\subsection{Comparison with predictions from simulation}

\begin{figure}[ht]
    \centering
    \includegraphics[width=8.5cm]{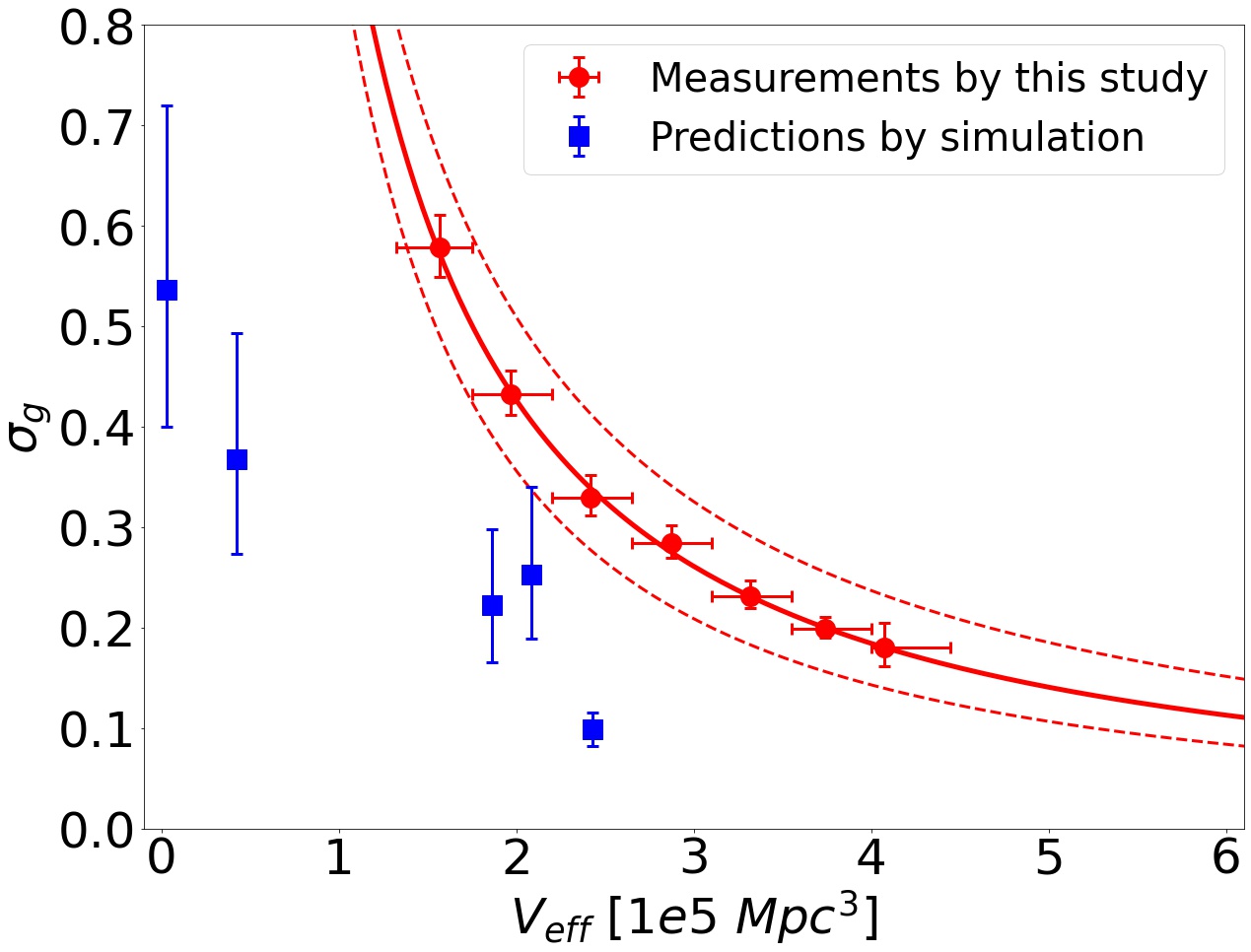}
    \caption{Comparison between $\sigma_g$ measured by this study (red dots and line) and predictions from simulations (blue squares). The red dots are measurements taken by altering the effective volume covered by each pointing. The red solid line is the best-fit power law described in section 4.2, and the red dashed lines correspond to the 1-$\sigma$ error range of the fitting parameters. The blue squares are calculated using the approach by \cite{Moster_etal_2010}.}
    \label{fig:CV_comp}
\end{figure}

In previous studies investigating Ly$\alpha$ LFs \cite[e.g.][]{Ouchi_etal_2008,Konno_etal_2016,Robertson_2010a}, cosmic variance is estimated from predictions of the $\Lambda$CDM model and N-body simulations. 
Using this approach, \cite{Moster_etal_2010} have derived a model to calculate $\sigma_g$ for five surveys (UDF, GOODS, GEMS, AEGIS, COSMOS), as a function of mean redshift $\bar{z}$, redshift bin size $\Delta z$, and the stellar mass of the galaxy population $m_*$. 
We apply their model with $\bar{z}=2.2$, $\Delta z=0.08$ (consistent with our data), and $8.5<log(m_*/M_{\odot})<9.0$ (typical stellar mass for LAEs, see e.g. \citealt{Ouchi_etal_2020}), and obtain the theoretical values of $\sigma_g$ predicted for 5 surveys at $z=2.16-2.24$. 
We over-plot these $\sigma_g$ against their corresponding survey volume as blue squares in figure \ref{fig:CV_comp}. 
Since the corresponding survey volume of the COSMOS survey is $\sim 18.8 \times 10^5 Mpc^3$, which exceeds the range of our measurement, we exclude this data point. 
We also over-plot the cosmic variance adopted by \cite{Konno_etal_2016} as a blue square in figure \ref{fig:CV_comp}, which is also estimated using the same method.

Measurements on $\sigma_g$ taken by our study and the fitting model described in section 4.2 are plotted as red dots and lines in figure \ref{fig:CV_comp}. 
It is shown that our measurements of $\sigma_g$ are significantly larger than the theoretical values (about 2-4 times). 
This suggests that the cosmic variance of LAEs might be different from that of general star-forming galaxies. 
This conclusion is consistent with \cite{Ito_etal_2021}'s results, which showed that the bias factor of LAEs might be strongly influenced by the HI distribution, and thus might be different from that of general galaxies.

Our results also suggest that the cosmic variances used by previous Ly$\alpha$ LF studies are likely underestimated. 
A possible cause is that previous approaches use correlation functions \cite[e.g.][]{Robertson_2010b} to estimate the galaxy bias, which only depends on the coordinates of detected galaxies, and then uses the galaxy bias and $\sigma_{\rm DM}$ to calculate $\sigma_g$. 
On the other hand, our measurements are carried out by integrating the LFs which take Ly$\alpha$ luminosity and completeness corrections into account. 
As a result, introducing Ly$\alpha$ luminosity and completeness correction may cause a higher cosmic variance that is likely more realistic for LF studies.

The underestimated cosmic variance can also explain the inconsistencies on LF fittings between different narrowband imaging surveys \cite[See e.g.][]{Konno_etal_2016,Sobral_etal_2017}. 
For instance, \cite{Konno_etal_2016} adopt a value of $0.099 \pm{0.017}$ for $\sigma_g$ using predictions by simulation and $\Lambda$CDM, while our model gives a value of $0.335_{-0.061}^{+0.074}$ for $\sigma_g$ under the same survey volume, which is significantly larger. 
This underestimation can further enlarge the error range of \cite{Konno_etal_2016}'s LFs, and thus provide another reason for the offset between \cite{Konno_etal_2016}'s LFs and ours presented in figure \ref{fig:LFlits}. 
Similar arguments can be applied to other LF studies, and thus explain the inconsistencies in the LF fittings by different surveys. 
Although, given that the cosmic variances are only added to the error bars during LF calculation, this underestimation will likely leave the major conclusions unchanged, only enlarging their error ranges.

\section{SUMMARY}
We have observed 8 overdense fields targeting the MAMMOTH candidates, with a total survey area of $\sim$ 11.63 deg$^2$. 
We have carried out a narrowband imaging survey using HSC with two narrowband filters NB387 and NB400 and investigated the Ly$\alpha$ LFs of LAEs selected using the narrowband color excess. 
The results are summarised below.
\begin{enumerate}
\item We fit a Schechter function to our LFs by having the faint-end slope $\alpha$ fixed at -1.75, which is adopted from \cite{Konno_etal_2016}. 
The best-fit values for the other two Schechter parameters are $L_{Ly\alpha}^{*}=5.18_{-0.40}^{+0.43}$ $ \times 10^{42}$ erg s$^{-1}$ and $\phi_{Ly\alpha}^{*}=4.87_{-0.55}^{+0.54}$ $ \times 10^{-4}$ Mpc$^{-3}$ for the overall data. 
Our results show a slightly lower $\phi_{Ly\alpha}^*$ but consistent $L_{Ly\alpha}^*$ compared to \cite{Konno_etal_2016} and \cite{Sobral_etal_2017}. 

\item Using the Ly$\alpha$ LFs of 200 LAE subsamples per field within the pointings created, we have investigated the field-to-field variation that arises from the cosmic variance. 
After clipping out the MAMMOTH pre-selected regions to reduce the bias, we create circular pointings on the field images with a fixed volume. 
We then calculate the LAE number densities for these pointings, by integrating the Ly$\alpha$ LFs with an integration limit of log$L_{\rm lim}$ = 41.41 erg s$^{-1}$.
We fit a log-normal function to the resulting distribution and compute $\sigma_g$. 

\item We have investigated the volume dependence of cosmic variance and obtained a relation between $\sigma_g$ and effective survey volumes for $z\sim$ 2.2 LAEs. 
We assume a simple power-law of $\sigma_g = k \times (\frac{V_{\rm eff}}{10^5 {\rm Mpc}^3})^{\beta}$ and obtain a value of $-1.209_{-0.106}^{+0.106}$ for $\beta$ and $0.986_{-0.100}^{+0.108}$ for k.

\item We compare our measurements of $\sigma_g$ with values predicted by simulations and find that our values are significantly larger. 
This suggests that previous Ly$\alpha$ LF studies may underestimate the cosmic variance, which can explain the different Ly$\alpha$ LFs between different surveys. 
Our results also imply that the cosmic variance of LAEs might be different from that of general star-forming galaxies.

\end{enumerate}

The Hyper Suprime-Cam (HSC) collaboration includes the astronomical communities of Japan and Taiwan, and Princeton University. The HSC instrumentation and software were developed by the National Astronomical Observatory of Japan (NAOJ), the Kavli Institute for the Physics and Mathematics of the Universe (Kavli IPMU), the University of Tokyo, the High Energy Accelerator Research Organization (KEK), the Academia Sinica Institute for Astronomy and Astrophysics in Taiwan (ASIAA), and Princeton University. Funding was contributed by the FIRST program from the Japanese Cabinet Office, the Ministry of Education, Culture, Sports, Science and Technology (MEXT), the Japan Society for the Promotion of Science (JSPS), Japan Science and Technology Agency (JST), the Toray Science Foundation, NAOJ, Kavli IPMU, KEK, ASIAA, and Princeton University. 

This paper makes use of software developed for the Large Synoptic Survey Telescope. We thank the LSST Project for making their code available as free software at  http://dm.lsst.org. 

The Pan-STARRS1 Surveys (PS1) have been made possible through contributions of the Institute for Astronomy, the University of Hawaii, the Pan-STARRS Project Office, the Max-Planck Society and its participating institutes, the Max Planck Institute for Astronomy, Heidelberg and the Max Planck Institute for Extraterrestrial Physics, Garching, The Johns Hopkins University, Durham University, the University of Edinburgh, Queen’s University Belfast, the Harvard-Smithsonian Center for Astrophysics, the Las Cumbres Observatory Global Telescope Network Incorporated, the National Central University of Taiwan, the Space Telescope Science Institute, the National Aeronautics and Space Administration under Grant No. NNX08AR22G issued through the Planetary Science Division of the NASA Science Mission Directorate, the National Science Foundation under Grant No. AST-1238877, the University of Maryland, and Eotvos Lorand University (ELTE) and the Los Alamos National Laboratory. 

Based in part on data collected at the Subaru Telescope and retrieved from the HSC data archive system, which is operated by Subaru Telescope and Astronomy Data Center at National Astronomical Observatory of Japan. 

The authors wish to recognize and acknowledge the very significant cultural role and reverence that the summit of Maunakea has always had within the indigenous Hawaiian community.  We are most fortunate to have the opportunity to conduct observations from this mountain. 

This paper includes data gathered with the 6.5 meter Magellan Telescopes located at Las Campanas Observatory, Chile. 

\bibliography{Reference}

\end{document}